\documentclass[acus]{JAC2003}

\usepackage{relsize}
\def\babar{\mbox{\slshape B\kern-0.1em{\smaller A}\kern-0.1em
    B\kern-0.1em{\smaller A\kern-0.2em R}}}

\def\epem       {\ensuremath{e^+e^-}\xspace}
\def\mpmm       {\ensuremath{\mu^+\mu^-}\xspace}
\def\pep2{PEP-II}
\def\BF{$B$ Factory}
\def\abf {asymmetric \BF}
\def\splumi {\ensuremath{{\cal L}_{\mathrm sp}}}

\RequirePackage{xspace}
\newcommand\slacpub  [1]{\rm SLAC-PUB-#1}

\newcommand{\tev}{\ensuremath{\mathrm{\,Te\kern -0.1em V}}\xspace}
\newcommand{\gev}{\ensuremath{\mathrm{\,Ge\kern -0.1em V}}\xspace}
\newcommand{\mev}{\ensuremath{\mathrm{\,Me\kern -0.1em V}}\xspace}
\newcommand{\kev}{\ensuremath{\mathrm{\,ke\kern -0.1em V}}\xspace}
\newcommand{\ev}{\ensuremath{\mathrm{\,e\kern -0.1em V}}\xspace}
\newcommand{\gevc}{\ensuremath{{\mathrm{\,Ge\kern -0.1em V\!/}c}}\xspace}
\newcommand{\mevc}{\ensuremath{{\mathrm{\,Me\kern -0.1em V\!/}c}}\xspace}
\newcommand{\gevcc}{\ensuremath{{\mathrm{\,Ge\kern -0.1em V\!/}c^2}}\xspace}
\newcommand{\mevcc}{\ensuremath{{\mathrm{\,Me\kern -0.1em V\!/}c^2}}\xspace}

\newcommand{\um}{\ensuremath{\mathrm{\mu m}}}
\newcommand{\nmrad}{\ensuremath{\mathrm{nm}}-\ensuremath{\mathrm{rad}}}
\newcommand{\mrad}{\ensuremath{\mathrm{mrad}}}

\newcommand{\nimBaseA}       {Nucl.\ Instr.\ Methods Phys.\ Res., Sect.\ 
A\xspace}

\newcommand{\nima}      [1]  {\nimBaseA~{\bf #1}}

\usepackage{graphicx}

\begin{document}
\title{COMBINED PHASE SPACE CHARACTERIZATION AT THE PEP-II IP USING 
       SINGLE-BEAM AND LUMINOUS-REGION MEASUREMENTS
\thanks{Work supported in part by DOE Contract DE-AC02-76SF00515}
\thanks{\slacpub{11904}}
}

\author{A. Bevan\thanks{bevan@slac.stanford.edu}, Queen Mary University of 
London, London, UK\\
        W. Kozanecki, DAPNIA-SPP, CEA-Saclay, F91191 Gif-sur-Yvette, France\\ 
        B. Viaud, Universit\'e de Montr\'eal, Montr\'eal, Qu\'ebec, Canada H3C 
3J7\\
        Y. Cai, A. Fisher, C. O'Grady, J. Thompson, M. Weaver, SLAC, Stanford 
CA 94309, U.S.A.\\
}

\maketitle

\begin{abstract}
We present a novel method to characterize the $e^\pm$ phase space at the 
IP of the SLAC B-factory, that combines single-beam measurements with a 
detailed mapping of luminous-region observables. Transverse spot sizes 
are determined in the two rings with synchrotron-light monitors and 
extrapolated to the IP using measured lattice functions. The specific 
luminosity, which is proportional to the inverse product of the overlap IP beam 
sizes, is continuously monitored using radiative--Bhabha events. The spatial 
variation of the luminosity and of the transverse-boost distribution of the 
colliding $e^{\pm}$, are measured using $e^+ e^- \to \mu^+ \mu^-$ events 
reconstructed in the \babar\ detector. The combination of these measurements 
provide constraints on the emittances, horizontal and vertical spot sizes, 
angular divergences and $\beta$ functions of both beams at the IP during 
physics data-taking. Preliminary results of this combined spot-size analysis 
are confronted with independent measurements of IP $\beta$-functions and 
overlap IP beam sizes at low beam current.

\end{abstract}

\section{INTRODUCTION}

The \babar\ detector~\cite{babar_nim} is located at the interaction point (IP) 
of the \pep2\ \abf~\cite{pep2}, where 3.1 \gev\ positrons from the low-energy 
ring (LER) collide with 9.0 \gev\ electrons from the high-energy ring (HER).  
The LER has two beam profile monitors: a visible synchrotron-light monitor 
($SLM_L$)~\cite{slmdesign} in a high-coupling region, and an X-ray monitor 
($SXM_L$)~\cite{sxmdesign} at a separate location. The HER is equipped with one 
$SLM_H$. In both rings, the vertical beam size at the SLM is measured using a 
companion interferometer~\cite{interferometer}. The \babar\ tracking system is 
used to measure the three-dimensional distribution of $\epem\to\mpmm$ 
vertices~\cite{bib:L3d} and the transverse-boost distribution~\cite{boost} of 
the muon pairs. This paper describes a first attempt at characterizing the 
phase space of the colliding beams by combining all the available information.

\par
\begin{figure}[!ht]
\centering
 \rotatebox{0}{\includegraphics[width=8.0cm]{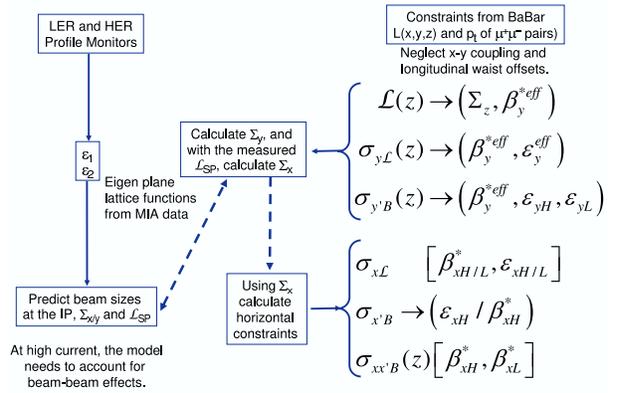}}
\caption{Schematic of possible inputs to a combined IP spot-size analysis.  
Variables in parentheses can be directly extracted from the parent observables; 
variables within square brackets are constrained by, but cannot be 
unambiguously determined from, the indicated distributions.
}\label{fig:outline}
\end{figure}

The strategy is outlined in Fig.~\ref{fig:outline}. In each ring, 
profile-monitor data are combined with measured lattice functions to extract 
the eigenmode emittances $\epsilon_{1,2}$ and predict the $e^{\pm}$ IP spot 
sizes. The z-dependence of the luminosity ${\cal L}$, of the luminous size 
$\sigma_{(x/y)_{\cal L}}$ and of the boost angular divergence 
$\sigma_{(x'/y')_B}$ allow the determination, under high-luminosity conditions, 
of the overlap bunch length $\Sigma_z$ and of the vertical emittances 
$\epsilon_{yH/L}$ and effective IP $\beta$-function $\beta_y^{*eff}$. Together 
with the measured specific luminosity $\splumi$, they also provide constraints 
on the horizontal emittances $\epsilon_{xH/L}$ and $\beta$ functions 
$\beta^*_{xH/L}$. Each set of observables 
(profile monitors, luminous-region data) 
offers a nearly complete description of the IP phase space. Comparing results 
for overlapping parameters is used to validate the techniques or identify 
inconsistencies, and combining all measurements should eventually yield a 
complete, and partially constrained, description.

%%  In this paper, we first discuss the method used to calculate the IP phase 
%%  space of the two beams using measured spot sizes at the profile monitors.
%%  We then describe results of measurements made using data recorded by 
%% \babar\ and finally discuss strategies for combining the available 
%%  information.

\section{USE OF BEAM-PROFILE MONITORS}

\begin{table*}[!ht]
\begin{center}
\caption{Eigenemittances (\nmrad) as inferred from single-beam profile-monitor 
data at low current, and as predicted by a simulation assuming the same lattice 
functions. Numbers in parentheses reflect higher-current measurements.
}
\footnotesize
\begin{tabular}{l|cc|cc}\hline
%%  \begin{tabular}{l|cccc}\hline
\textbf{Input} &  \textbf{LER \boldmath $\epsilon_1$} &   \textbf{LER \boldmath 
$\epsilon_2$ }  &  \textbf{HER \boldmath $\epsilon_1$ } &   \textbf{HER 
\boldmath $\epsilon_2$ } \\ \hline
$SXM ~only: \sigma_x, \sigma_y$   &  $49 \pm 13$   ($29^{+9}_{-8}$)      &   
$2.3 \pm 0.5$ ($4\pm 0.8$)           & $-$ & $-$\\
$SLM ~only: \sigma_x, \sigma_y$   &  $34 \pm 8$ ($35 \pm 8$)             &   $-
37\pm 9$ ($-43^{+9}_{-7}$)         & $130 \pm 38$  ($190\pm 50$)          &   
$1.6 \pm 0.3$ ($2.8^{+0.5}_{-0.8}$) \\
$SLM+SXM: \sigma_x$ only   &  $49^{+17}_{-12}$ ($36^{+11}_{-10}$) & $-2.7 \pm 
1.0$   ($0.7\pm 5$)& $-$ & $-$\\
$SLM+SXM: \sigma_x, \sigma_y$, $\psi$ & $33.6 \pm 7.4$ & $2.5 \pm 1.2$ & $-$ & 
$-$\\
\hline
LEGO simulation             &  32.6 & 1.13  & 50 & 0.36 \\ \hline
\end{tabular}
\label{tbl:characterisationdataprofileemittances}
\end{center}
\end{table*}

The projected beam sizes $\sigma_x$ and $\sigma_y$ and the tilt angle $\psi$ of 
the transverse profile are measured at each of the three spot-size monitors. 
Lattice properties are measured by resonant excitation, one ring at a time in 
single-bunch mode. 
The beam-position monitor data are analyzed using a model-independent technique 
(MIA)~\cite{MIA}, and fitted in the context of the LEGO 
package~\cite{yunhaimodel} to produce a set of fully-coupled lattice functions 
using the formalism of Ref.~\cite{oneturnmatrix}. The same procedure 
predicts the $e^{\pm}$ eigenemittances in the absence of beam-beam 
interactions.

\par
Using, at each profile monitor, the measured one-turn matrix extracted above, 
one can express the measured beam size in terms of two 
(unknown) 
eigenmode emittances $\epsilon_{1,2}$ and of ten 
%%  fitted
lattice parameters:
{
\small
\begin{eqnarray}
 \sigma_x^2=\beta_1 \epsilon_1 g^2 + (\beta_2 w_{22}^2 + 2 \alpha_2 
w_{22}w_{12} + \gamma_2 w_{12}^2)\epsilon_2 + \sigma_{\eta_x}^2,\\ 
 \sigma_y^2=(\beta_1 w_{11}^2 - 2 \alpha_1 w_{11}w_{12} + \gamma_1 
w_{12}^2)\epsilon_1 + \beta_2 \epsilon_2 g^2 + \sigma_{\eta_y}^2, 
\end{eqnarray}
}where $\beta_i$ are the eigenmode $\beta$ functions at the source point, 
$\alpha_i = -\beta_i^\prime/2$, $\gamma_i = (1+\alpha_i^2)/\beta_i$,
$w$ is a $2\times 2$ quasi-symplectic matrix describing coupling between $x$ and $y$, 
$g^2 = 1-\det(w)$, and $\sigma_{\eta_{x,y}} = \eta_{x,y} \Delta p/p$ are the 
dispersive contributions to the projected beam sizes.  $x$-$y$ coupling also 
manifests itself by a tilted profile-monitor image:
{
\small
\begin{eqnarray}
 \sigma_{xy}&=&g[(\beta_2 w_{22} + \alpha_2 w_{12})\epsilon_2 - (\beta_1 w_{11}  
\nonumber \\
              && - \alpha_1 w_{12})\epsilon_1 ] + 
\sigma_{\eta_x}\sigma_{\eta_y}, \\
 \tan (2\psi) &=& 2\sigma_{xy} / ( \sigma_x^2 - \sigma_y^2).
\end{eqnarray}
}

\par
The eigenemittance being an invariant, it should not depend on where around the 
ring it is measured. In order to verify the consistency of the fitted lattice 
functions with the profile-monitor data, dedicated beam-size measurements were 
performed at low bunch currents in both single- and colliding-beam 
configurations.
Table~\ref{tbl:characterisationdataprofileemittances} summarizes the 
eigenemittances inferred using various combinations of input measurements. The 
errors quoted assume a $\pm$10\% uncertainty on each measured spot size and 
$\pm 2^o$ on the tilt angles. The results for $\epsilon_{1,L}$ are consistent 
within errors. The unphysical values of $\epsilon_{2,L}$ are probably due to 
the fact that the $SLM_L$ vertical beam size is dominated by $\epsilon_1$, so 
that small measurement or lattice-function errors have a disproportionate 
impact on $\epsilon_2$. In contrast, the nominal coupling is zero at $SXM_L$ 
and its $\epsilon_2$ measurement intrinsically more reliable. The most robust 
LER result is provided by a constrained SVD fit that combines all the experimental
information available at $SLM_L$ and $SXM_L$. Except for $\epsilon_{1,H}$ which 
is more than a factor of two larger than expected, the eigenemittances inferred 
from the profile monitors are acceptably consistent with those predicted by the 
simulation.

\par
An additional check is provided by collision data. The $e^-$ and $e^+$ IP spot 
sizes are estimated by combining profile-monitor results with the values of the 
lattice parameters extrapolated to the IP. The resulting predicted overlap beam 
sizes $\Sigma^2_{pred,j} = \sigma_{LER,j}^2 + \sigma_{HER,j}^2$ ($j=major, 
minor$) and tilts can then be compared to those measured by an overconstrained 
set of beam-beam scans~\cite{oneturnmatrix}.
%%  after correcting $\Sigma_{minor}$ for the hourglass 
%%  effect~\cite{venturini}. 
This procedure is only feasible at low bunch current, when the beam-beam 
parameters are small enough not to distort the one-turn matrix or render such 
scans impractical. 
The results are summarised in Table~\ref{tbl:capsigmaresults}. The discrepancy 
between the measured and predicted values of $\Sigma_{major}$ is due to the 
unphysically large value of $\epsilon_{1,H}$. If instead one estimates 
$\epsilon_{1,H}$ using $\Sigma_{msrd, major}$ and $\sigma_{LER, major}$, one 
obtains $\epsilon_{1,H}=28^{+7}_{-9}$\,\nmrad,
which is closer to expectations than the value inferred from $SLM_H$. 
Similarly, the specific luminosity $\splumi$ inferred from $\Sigma_{msrd}$ 
($\splumi=4.0$~$\mu b^{-1} s^{-1} bunch^{-1}(mA/bunch)^{-2}$) is consistent 
with that directly measured by the radiative-Bhabha monitor ($\splumi=4.3$), 
and significantly larger than that predicted using $\Sigma_{pred}$ 
($\splumi=2.5^{+0.4}_{-0.5}$). The primary mirror for $SLM_H$ (a water-cooled 
mirror in vacuum that withstands a large heat load) is known to have mechanical 
stresses that may distort the image somewhat.

%%  A similar conclusion is reached when comparing the specific luminosity 
%%  $\splumi$ predicted using 
%%  $\Sigma_{pred}$ ($2.5^{+0.4}_{-0.5}$
%%  ~$\mu b^{-1} s^{-1} bunch^{-1}(mA/bunch)^{-2}$) with that inferred from
%%   $\Sigma_{msrd}$ ($\splumi=4.0$), or that directly measured by the 
%%  radiative-Bhabha monitor ($\splumi=4.3$).
\vspace{-0.3cm}
\begin{table}[!ht]
\begin{center}
\caption{$e^{\pm}$ IP beam sizes and tilts predicted using eigenemittances 
inferred from $SLM_L$, $SXM_L$ and $SLM_H$ data and fitted lattice functions. 
The predicted values ($\Sigma_{pred}$) are compared to those directly measured 
($\Sigma_{msrd}$).  
}
\footnotesize
\begin{tabular}{l|ccc}\hline
\textbf{Input}     &  \textbf{\boldmath Major axis (\um)} &  \textbf{\boldmath 
Minor axis (\um)}  &  \textbf{\boldmath $\psi$ (\mrad)} \\ \hline
$\sigma_{LER}$     & $121 \pm 16$         & $4.7^{+0.4}_{-0.7}$   & -11.3 \\ 
$\sigma_{HER}$     & $244\pm 35$          & $4.3 \pm 0.4$         & -16.7 \\
$\Sigma_{pred}$  & $271^{+39}_{-10}$    & $6.8\pm 0.7$          & -15.7 \\
\hline
$\Sigma_{msrd}$    & 169.4                & 6.7                   & -7.3  \\
\end{tabular}
\label{tbl:capsigmaresults}
\end{center}
\end{table}
\vspace{-0.3cm}

\par
Spot sizes were also recorded at somewhat higher currents. In the absence of 
collisions and with an $e^+$ current far below any potential $e^-$-cloud 
threshold, the emittances should remain the same. The observed values are 
comparable to the low-current results 
(Table~\ref{tbl:characterisationdataprofileemittances}). The increase in 
$\epsilon_{2,L}$ may be caused by thermal orbit distortions.
The already large horizontal HER spot size appears to grow with current.

\par
Data recorded with high-current colliding beams exhibit a sizeable increase in 
the vertical $SXM_L$ and in both $SLM_L$ spot sizes, which is qualitatively 
consistent with the dynamic-$\beta$ effect and with beam-beam vertical blowup. 
But the observed spot sizes result in inconsistent emittance estimates, 
presumably highlighting the fact that the additional focusing caused by the 
beam-beam interaction must imperatively be taken into account in the 
one-turn matrix.

\section{LUMINOUS-REGION ANALYSIS}

The size of the luminous ellipsoid~\cite{bib:L3d} and the transverse-boost 
distribution~\cite{boost} of the colliding electron and positron are measured 
using $\epem\to\mpmm$ events reconstructed in the \babar\ detector. The spatial 
variation of these observables is determined by the emittances, IP 
$\beta$-functions and waist locations of the colliding beams.

\par
\begin{figure}[htb]
\centering
\includegraphics[width=2.3in]{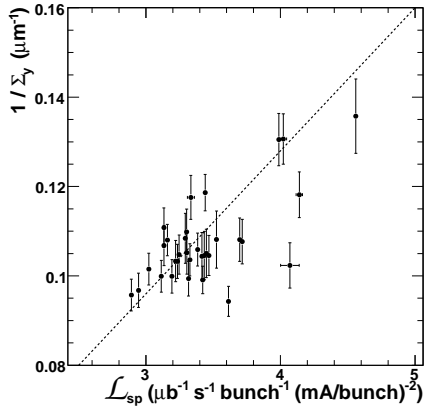}
\caption{Correlation between $1/\Sigma_y$ (boost method) and $\splumi$ measured 
by the Bhabha luminosity monitor.
}
\label{fig:CapsigyvLsp}
\end{figure}

In the horizontal plane, $\beta^*_{xH/L}$ is much larger than the bunch lengths 
$\sigma_{zH/L}$, so the horizontal beam parameters are only weakly apparent in 
the $z$-dependence of luminous-region variables. 
But in the vertical plane, $\beta^*_{yH/L} \sim \sigma_{zH/L}$, resulting in an 
observable $z$-dependence of the luminosity, of the angular spread of the boost 
direction and of the vertical size of the luminous region. Fitting an effective 
IP $\beta$-function to the first two observables and neglecting $x$-$y$ 
coupling yields similar results, in the range of 12--16\,mm~\cite{bib:L3d}; 
fits to $\sigma_{y_{\cal L}}\left(z\right)$ yield somewhat higher values, but 
with larger systematic uncertainties. An effective vertical emittance can then 
be extracted from $\sigma_{y_{\cal L}}\left(z=0\right)$; the more powerful 
boost technique allows to determine both the HER and LER emittances, yielding 
$\epsilon_{yH,L}\sim$2.5--9\,\nmrad\ (again under the no-coupling 
assumption)~\cite{boost}. Combining the emittance and $\beta^*$ results from 
the boost measurement yields estimates of $\Sigma_y$ in the range of 
7--10\,$\mu$m, that displays the expected anticorrelation with the specific 
luminosity (Fig.~\ref{fig:CapsigyvLsp}). Because 
$\splumi \sim 1/\Sigma_x \Sigma_y$, the slope of this correlation provides a 
measurement of $\Sigma_x$.

\par
\begin{figure}[htb]
\centering
\includegraphics[width=3.2in]{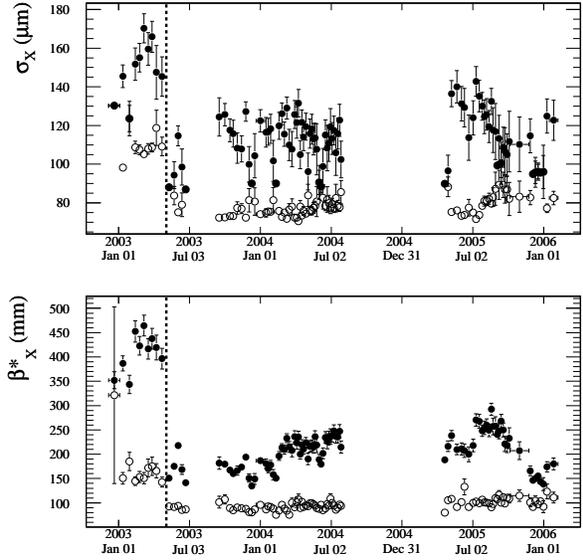}
\caption{History of horizontal IP spot sizes (top) and $\beta$ functions 
(bottom) in the LER (open circles) and the HER (black dots), extracted from 
luminous-region observables measured by \babar. The dotted line indicates the 
time when both $x$ tunes were moved close to the half-integer, resulting in a 
sizeable luminosity improvement.}
\label{fig:babarx}
\end{figure}

Further combining $\Sigma_x$ with the horizontal luminous size $\sigma_{x_{\cal 
L}}$~\cite{bib:L3d} determines both $e^+$ and $e^-$ horizontal IP spot sizes 
(Fig.~\ref{fig:babarx}). These are then finally combined with the 
horizontal-boost angular spread $\sigma_{x'_B}$ and the horizontal 
angle-position correlation $(\delta x'/\delta x)_B$~\cite{boost} to extract 
$\beta^*_{xH,L}$. The reduction in horizontal spot size caused by the 
dynamic-$\beta$ effect is strikingly apparent. The value of $\beta^*_{xH}$ 
extracted from \babar\ data before the move to the half-integer, is consistent 
with low-current phase advance measurements recorded at that time. In the same 
period however, the corresponding $\beta^*_{xL}$ result is significantly 
smaller than the phase-advance measurements. In addition, the horizontal LER 
emittance implied by Fig.~\ref{fig:babarx} is surprisingly large, possibly 
signaling an inconsistency in this preliminary analysis.

\section{CONCLUSION}

Confronting IP beam sizes extrapolated from the profile monitors with a 
detailed phase-space characterization based on luminous-region observables 
would provide invaluable consistency checks, as well as restrict systematic 
uncertainties. Two main ingredients are required to this effect. First, 
beam-beam focusing needs to be taken into account in the one-turn matrix of 
Ref.~\cite{oneturnmatrix}; guidance can be provided here by beam-beam 
simulations. Second, $x$-$y$ coupling at the IP needs to be incorporated in the 
luminous-region analyses, as its impact has recently been 
shown~\cite{bib:L3d,boost} to be significant.


\begin{thebibliography}{9}   % Use for  1-9  references

\bibitem{babar_nim}
\babar\ Collaboration, B.\ Aubert {\em et al.}, \nima{479}, 1 (2002).

\bibitem{pep2}
J. Seeman {\em et al.}, EPAC-2006-MOPLS045.

\bibitem{slmdesign}
A. Fisher {\em et al.}, AIP Conference Proceedings 451, Woodbury, NY:
American Institute of Physics 95-109 (1998).

\bibitem{sxmdesign}
A. Fisher {\em et al.}, \slacpub{10528} (2004).

\bibitem{interferometer}
A. Fisher {\em et al.}, \slacpub{8867} (2001).

\bibitem{bib:L3d}
J.~Thompson {\em et al.}, \slacpub{11222} (2005); B. Viaud  {\em et al.}, 
\slacpub{11900} (2006).

\bibitem{boost}
M.~Weaver {\em et al.}, \slacpub{11906} (2006).

\bibitem{MIA}
Y. Yan {\em et al.}, \slacpub{11209} (2005).

\bibitem{yunhaimodel}
Y. Cai {\em et al.}, \slacpub{7642} (1997); Y. Cai {\em et al.}, 
EPAC-2006-MOPLS052.

\bibitem{oneturnmatrix}
Y. Cai, \slacpub{8479} (2000).

%%  \bibitem{venturini}
%%  M.~Venturini and W.~Kozanecki, \slacpub{8699} (2000).

\end{thebibliography}
\end{document}